\documentclass[12pt,a4paper]{article}
\usepackage{amsfonts}
\usepackage{amssymb}
\usepackage{amsmath}
\usepackage{latexsym}
\usepackage{epsf,epsfig,rotating}

\begin{document}

\begin{center}
{\Large \bf Gravity in Brans-Dicke theory with
Born-Infeld scalar field and the Pioneer anomaly} \\

\vspace{4mm}

M.N.~Smolyakov\\
\vspace{0.5cm} Skobeltsyn Institute of Nuclear Physics, Moscow
State University,
\\ 119991, Moscow, Russia\\
\end{center}

\begin{abstract}
In this paper we discuss a model which can be considered as a
generalization of the well-known scalar-tensor Brans-Dicke theory.
This model possesses an interesting feature: due to Born-Infeld
type non-linearity of the scalar field the properties of the
interaction between two test bodies depend significantly on their
masses. It is shown that the model can be interesting in view of
the Pioneer 10, 11 spacecraft anomaly.
\end{abstract}

\section{Introduction and setup}
One of the most known scalar-tensor theories of gravity is the
Brans-Dicke theory \cite{MTW,Weinberg}. It describes the scalar
field non-minimally coupled to gravity with the action
\begin{eqnarray}\label{act-BD-1}
S=\int d^{4}x\sqrt{-g}\left[\varphi
R-\tilde\omega\frac{g^{\mu\nu}\partial_{\mu}\varphi\partial_{\nu}\varphi}{\varphi}+L_{matter}\right],
\end{eqnarray}
where $\tilde\omega$ is the Brans-Dicke parameter and $L_{matter}$
is the Lagrangian of matter. In the limit $\tilde\omega\to\infty$
the theory goes to the standard General Relativity. This theory is
very well examined, the present days gravitational experiments set
stringent limits on possible values of the Brans-Dicke parameter
$\tilde\omega$ \cite{Will:2005va,Chiba:2003ir}.

In this paper we consider a generalization of this theory based on
the use of the Born-Infeld scalar field. This field itself was
widely discussed in the literature, see, for example,
\cite{BC}--\cite{Gibbons:2001gy} and references therein. We will
show that such a highly non-linear covariant theory possesses very
interesting features and, in principle, it can account for the
anomalous acceleration of Pioneer~10 and Pioneer~11 spacecraft
\cite{Anderson:2001sg,Toth:2006qb}, leaving the planets of the
Solar System devoid of such extra constant acceleration which is
excluded by observations \cite{Sanders:2006yf}.

To this end let us consider the following four-dimensional action
describing Born-Infeld scalar field non-minimally coupled to
gravity
\begin{eqnarray}\label{act-BD}
S=\int d^{4}x\sqrt{-g}\left[\varphi
R+f\sqrt{1-\frac{\omega}{f}\frac{g^{\mu\nu}\partial_{\mu}\varphi\partial_{\nu}\varphi}{\varphi}}-
f+L_{matter}\right]
\end{eqnarray}
with $f>0$. The action of the scalar field has a non-standard
form, but in the limit $f\to\infty$
$$f\sqrt{1-\frac{\omega}{f}\frac{g^{\mu\nu}\partial_{\mu}\varphi\partial_{\nu}\varphi}{\varphi}}\to
f-\frac{\omega}{2}\frac{g^{\mu\nu}\partial_{\mu}\varphi\partial_{\nu}\varphi}{\varphi},
$$
and (\ref{act-BD}) transforms into the well-know action of the
Brans-Dicke theory (\ref{act-BD-1}) with the Brans-Dicke parameter
$\tilde\omega=\frac{\omega}{2}$. The theory with action
(\ref{act-BD}) can be considered as a generalized Brans-Dicke
theory. The extra term $-f$ in (\ref{act-BD}) is added to preserve
the Minkowski background metric in the vacuum state.

The vacuum expectation value of the field $\varphi$ is supposed to
be $\varphi_{vac}=M_{Pl}^{2}$. Let us represent the scalar and the
gravitational fields as
\begin{equation}\label{repres}
\varphi=M_{Pl}^{2}+\frac{M_{Pl}}{\sqrt{\omega}}\phi,
\end{equation}
\begin{equation}\label{repres1}
g_{\mu\nu}=\eta_{\mu\nu}+\frac{1}{M_{Pl}}h_{\mu\nu},
\end{equation}
where $\phi_{vac}=0$, $g_{\mu\nu}^{vac}=\eta_{\mu\nu}$,
$\eta_{\mu\nu}=diag(-1,1,1,1)$ is the flat Minkowski metric, and
expand action (\ref{act-BD}) into series with respect to $\phi$
and $h_{\mu\nu}$. The "gravitational" part of the action
$$
\int d^{4}x\sqrt{-g}\varphi R
$$
can be represented as
\begin{eqnarray}\label{act-BD22}
\int
d^{4}x\left[M_{Pl}L^{(1)}[h_{\mu\nu}]+\frac{1}{\sqrt{\omega}}\phi
L^{(1)}[h_{\mu\nu}]+
\left(1+\frac{1}{M_{Pl}\sqrt{\omega}}\phi\right)L^{(2)}[h_{\mu\nu}]+...\right],
\end{eqnarray}
where $L^{(1)}[h_{\mu\nu}]$ is linear in $h_{\mu\nu}$,
$L^{(2)}[h_{\mu\nu}]$ is quadratic in $h_{\mu\nu}$ and so on. The
Born-Infeld part of the action
$$
\int
d^{4}x\sqrt{-g}f\sqrt{1-\frac{\omega}{f}\frac{g^{\mu\nu}\partial_{\mu}\varphi\partial_{\nu}\varphi}{\varphi}}
$$
can be rewritten as
\begin{eqnarray}\label{act-BD23}
\int
d^{4}xf\sqrt{1-\frac{1}{f}\left(\eta^{\mu\nu}-\frac{1}{M_{Pl}}h^{\mu\nu}+...\right)\partial_{\mu}\phi\partial_{\nu}\phi\left(1-\frac{1}{M_{Pl}\sqrt{\omega}}\phi+...\right)},
\end{eqnarray}
where we have omitted the expansion of $\sqrt{-g}$. Here and below
indices are raised by $\eta^{\mu\nu}$.

Now let us discuss formulas (\ref{act-BD22}) and (\ref{act-BD23}).
First, the term $L^{(1)}[h_{\mu\nu}]$ is simply
$L^{(1)}[h_{\mu\nu}]=\partial^{\mu}\partial^{\nu}h_{\mu\nu}-\partial^{\mu}\partial_{\mu}h$
(where $h=h_{\nu}^{\nu}$), which is a total derivative. Thus, the
term $M_{Pl}L^{(1)}[h_{\mu\nu}]$ vanishes from the action. Second,
since we suppose to work in the Newtonian approximation, we can
drop the term $\frac{1}{M_{Pl}}h^{\mu\nu}$ (as well as higher
corrections in $h_{\mu\nu}$) in comparison with $\eta^{\mu\nu}$.
For these reasons we can also drop the terms $L^{(n)}[h_{\mu\nu}]$
for $n>2$. As for the term $\frac{1}{M_{Pl}\sqrt{\omega}}\phi$, it
is not evident that it is much smaller than unity. Nevertheless,
let us suppose that $\frac{1}{M_{Pl}\sqrt{\omega}}\phi\ll 1$ and
drop the term $\frac{1}{M_{Pl}\sqrt{\omega}}\phi$ and the
subsequent terms in (\ref{act-BD23}), as well as the cubic term
$\frac{1}{M_{Pl}\sqrt{\omega}}\phi L^{(2)}[h_{\mu\nu}]$. Below we
will show that condition $\frac{1}{M_{Pl}\sqrt{\omega}}\phi\ll 1$
indeed holds. Note that we are not able to drop the quadratic term
$\frac{\phi}{\sqrt{\omega}}L^{(1)}[h_{\mu\nu}]$ because it ensures
the interaction of Born-Infeld scalar field with matter.

Thus we get
\begin{eqnarray}\label{act}
S_{eff}=\int d^{4}x\left(\frac{\phi}{\sqrt{\omega}}
L^{(1)}[h_{\mu\nu}]+ L^{(2)}[h_{\mu\nu}]+
f\sqrt{1-\frac{1}{f}\eta^{\mu\nu}\partial_{\mu}\phi\partial_{\nu}\phi}\,+\right.\\
\nonumber \left.\frac{1}{2M_{Pl}}h^{\mu\nu}t_{\mu\nu}\,\right),
\end{eqnarray}
where $t_{\mu\nu}$ is the energy-momentum tensor of matter and
\begin{eqnarray}
&&L^{(2)}[h_{\mu\nu}]=L_{FP}[h_{\mu\nu}]=\\
\nonumber
&&-\frac{1}{4}\left[\partial_{\rho}h_{\mu\nu}\partial^{\rho}h^{\mu\nu}-
\partial_{\rho}h\partial^{\rho}h+2\partial_{\mu}h^{\mu\nu}\partial_{\nu}h-2\partial_{\mu}h^{\mu\nu}
\partial^{\rho}h_{\rho\nu}\right]
\end{eqnarray}
is the standard Fierz-Pauli Lagrangian. It is convenient to
diagonalize action (\ref{act}) with the help of the standard
redefinition
\begin{equation}\label{subst2}
h_{\mu\nu}=b_{\mu\nu}-\frac{1}{\sqrt{\omega}}\eta_{\mu\nu}\phi.
\end{equation}
After some algebra we get
\begin{eqnarray}\label{Lagr}
L_{eff}=L_{FP}[b_{\mu\nu}]+f\sqrt{1-\frac{1}{f}\eta^{\mu\nu}\partial_{\mu}\phi\partial_{\nu}\phi}
-
\frac{3}{2\omega}\eta^{\mu\nu}\partial_{\mu}\phi\partial_{\nu}\phi+\\
\nonumber \frac{1}{2M_{Pl}}b^{\mu\nu}t_{\mu\nu}-
\frac{1}{2M_{Pl}\sqrt{\omega}}\phi t,
\end{eqnarray}
where $t=\eta^{\mu\nu}t_{\mu\nu}$. The extra term
$-\frac{3}{2\omega}\eta^{\mu\nu}\partial_{\mu}\phi\partial_{\nu}\phi$
in (\ref{act}) has appeared in the action after diagonalization.

The Born-Infeld part of action (\ref{Lagr}) has the standard form
of the Dirac-Born-Infeld (DBI) scalar field action, though the
standard DBI Lagrangian has a different origin. It is necessary to
note that we take $f>0$ (like in \cite{Mukhanov:2005bu}), contrary
to the case $f<0$, which is often discussed in the literature
(see, for example, \cite{BC}--\cite{Lu:2007vr},
\cite{Pavlovsky:2007pj}).

It should be also noted that we neglected the term $-f$ (see
action (\ref{act-BD})) while obtaining Lagrangian (\ref{Lagr}). We
will discuss this issue in the next section.

Lagrangian (\ref{Lagr}) allows one to examine the stability of the
model at least above the Minkowski background. To this end we
consider $t_{\mu\nu}=0$ and suppose that
$\eta^{\mu\nu}\partial_{\mu}\phi\partial_{\nu}\phi\ll f$.
Expanding the square root in (\ref{Lagr}) up to the linear term we
get a quadratic Lagrangian
\begin{eqnarray}\label{Lagr-2q}
L_{2}=L_{FP}[b_{\mu\nu}]-\frac{\omega+3}{2\omega}\eta^{\mu\nu}\partial_{\mu}\phi\partial_{\nu}\phi.
\end{eqnarray}
For $\omega\gg 1$ (this case will be discussed below) the kinetic
term of the scalar field $\phi$ has the proper sign, which leads
to the absence of ghosts in the theory. Indeed, for
$\eta^{\mu\nu}\partial_{\mu}\phi\partial_{\nu}\phi\ll f$
Lagrangian (\ref{Lagr}) describes the standard Brans-Dicke theory
in the Newtonian approximation, which is known to be stable.
Higher corrections to (\ref{Lagr-2q}) lead to an infinite tower of
self-interaction and interaction terms. Nevertheless, the solution
which will be discussed below corresponds to a deep non-linear
regime of the Born-Infeld part of the model where perturbation
theory does not work. It has appeared to be very difficult (maybe
even impossible) to examine the perturbations around this solution
analytically, one should make numerical analysis. Thus, the
question about the stability of the solution presented below has
no definite answer yet.

\section{Equations of motion and extra anomalous force}

The equations of motion following from Lagrangian (\ref{Lagr})
take the form
\begin{equation}\label{eq-gr-b}
\Box
b_{\mu\nu}=-\frac{1}{M_{Pl}}\left(t_{\mu\nu}-\frac{1}{2}\eta_{\mu\nu}t\right),
\end{equation}
\begin{equation}\label{eq1}
\partial_{\mu}\left(\frac{\eta^{\mu\nu}\partial_{\nu}\phi}{\sqrt{1-\frac{1}{f}\eta^{\rho\sigma}\partial_{\rho}\phi\partial_{\sigma}\phi}}
+\frac{3}{\omega}\eta^{\mu\nu}\partial_{\nu}\phi\right)=
\frac{1}{2M_{Pl}\sqrt{\omega}}t,
\end{equation}
where $\Box=\eta^{\mu\nu}\partial_{\mu}\partial_{\nu}$. We used de
Donder gauge condition
$\partial^{\mu}b_{\mu\nu}-\frac{1}{2}\partial_{\nu}b=0$ while
obtaining (\ref{eq-gr-b}).

We will be interested in the additional interaction between two
bodies caused by the scalar field $\phi$. As can be seen from
initial action (\ref{act-BD}), ordinary matter interacts only with
the metric, in this sense the weak equivalence principle is
fulfilled. Thus the gravitational force acting on a test body can
be easily obtained by considering geodesic motion, and in the
Newtonian limit we get the well-known formula
\begin{equation}
\ddot{\vec x}=\frac{1}{2M_{Pl}}\nabla h_{00}.
\end{equation}
If one considers a non-point-like source, this formula transforms
into a formula describing the force acting on the center of mass
of the test body
\begin{equation}
m\ddot{\vec R}=\frac{1}{2M_{Pl}}\int_{V}dV\rho(\vec x)\nabla
h_{00},
\end{equation}
where $\rho(\vec x)$ is the density of the body of volume $V$ such
that
$$\int_{V}\rho(\vec x) dV=m$$
and $\vec R$ is the vector pointing to the center of mass of the
body. Using (\ref{subst2}) we get the standard Newtonian force
(the contribution of $b_{\mu\nu}$), and an anomalous force
\begin{equation}\label{force}
{\vec F}_{anom}=\frac{1}{2M_{Pl}\sqrt{\omega}}\int_{V}dV\rho(\vec
x)\nabla\phi
\end{equation}
(the contribution of $\phi$). We will be interested in this
anomalous extra force.

Now let us turn to Eq. (\ref{eq1}) and consider the static case of
spherically symmetric bodies. We suppose that the energy-momentum
tensors of the test bodies have the form
$$t^{1,2}_{00}=\rho_{1,2}(\vec x),\quad t^{1,2}_{ij}=0.$$
Due to the spherical symmetry
\begin{eqnarray}\label{dens}
\rho_{2}(\vec x)=\rho(r),\quad r\le r_{*},\\ \label{dens1}
\rho_{2}(\vec x)=0,\quad r>r_{*},
\end{eqnarray}
where $r_{*}$ is the radius of the second body.

Let us denote $\eta^{ij}\partial_{j}\phi=\nabla\phi=\vec\phi$.
Then Eq. (\ref{eq1}) for the case of two test bodies takes the
form
\begin{equation}\label{eq2}
div\left(\frac{\vec\phi}{\sqrt{1-\frac{1}{f}({\vec\phi}{\vec\phi})}}+\frac{3}{\omega}\vec\phi\right)=
4\pi\left(-\frac{1}{8\pi \sqrt{\omega}M_{Pl}}\rho_{1}(\vec
x)-\frac{1}{8\pi\sqrt{\omega} M_{Pl}}\rho_{2}(\vec x)\right).
\end{equation}

\begin{figure}[h]
\centering
\includegraphics[width=12cm,height=3.84cm]{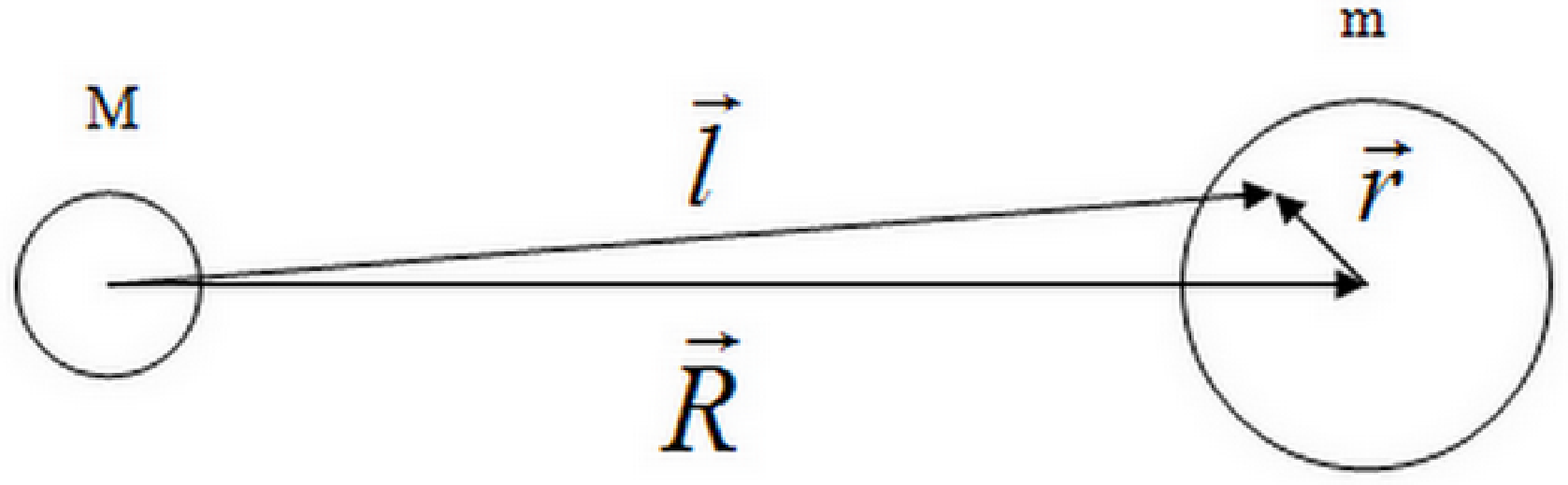}
\caption{Coordinate system used for calculations} \label{fig}
\end{figure}

Now we are ready to examine the force acting on the test body in
such a system. The coordinate system that will be used for
calculations is presented in Fig.~\ref{fig}. The force will be
calculated for the second body (the right body in Fig.~\ref{fig}).

The solution to Eq. (\ref{eq2}) inside the second body has the
form
\begin{equation}\label{eq3}
\frac{\vec\phi}{\sqrt{1-\frac{1}{f}({\vec\phi}{\vec\phi})}}+\frac{3}{\omega}\vec\phi=
-\frac{1}{8\pi\sqrt{\omega} M_{Pl}} \left(\frac{M\vec
l}{l^{3}}+\frac{m(r)\vec r}{r^{3}}\right),
\end{equation}
where $\vec l=\vec R+\vec r$, $r=\sqrt{\vec r^{2}}$, $M$ is the
mass of the first body,
$$m(r)=4\pi\int_{0}^{r} {\tilde r}^{2}\rho(\tilde r)d\tilde r$$
for $r<r_{*}$ and $m(r)|_{r\ge r_{*}}=m$ (see Fig.~\ref{fig}).

It is convenient to represent the parameter $f$ as
\begin{equation}\label{fexp}
f=\frac{M^{2}}{r_{M}^{4}(8\pi)^{2}\omega M_{Pl}^{2}},
\end{equation}
where $r_{M}$ is a parameter depending on $M$.

We suppose that $\omega\gg 1$. Since
$\frac{1}{\sqrt{1-\frac{1}{f}({\vec\phi}{\vec\phi})}}>1$ and
$3/\omega\ll 1$, we can look for a solution to Eq. (\ref{eq3})
using the perturbative approach. After some algebra one can get
\begin{equation}\label{eq3-0}
\vec\phi=-\frac{\sqrt{f}\left(\frac{M\vec l}{l^{3}}+\frac{m(r)\vec
r}{r^{3}}\right)}{\sqrt{\frac{M^{2}}{r_{M}^{4}}+\left(\frac{M\vec
l}{l^{3}}+\frac{m(r)\vec
r}{r^{3}}\right)^{2}}}\left[1-\frac{3}{\omega}\left(\frac{\frac{M^{2}}{r_{M}^{4}}}{\frac{M^{2}}{r_{M}^{4}}+
\left(\frac{M\vec l}{l^{3}}+\frac{m(r)\vec
r}{r^{3}}\right)^{2}}\right)^{\frac{3}{2}}\right].
\end{equation}
For the objects which will be discussed in the next section of the
paper the correction in (\ref{eq3-0}) $\sim 3/\omega$ appears to
be at least $\sim 10^{-3}$ or even smaller. Thus we can drop this
correction and use
\begin{equation}\label{gen-sol}
\vec\phi=-\left(\frac{M\vec l}{l^{3}}+\frac{m(r)\vec
r}{r^{3}}\right)\frac{\sqrt{f}}{\sqrt{\frac{M^{2}}{r_{M}^{4}}+\left(\frac{M\vec
l}{l^{3}}+\frac{m(r)\vec r}{r^{3}}\right)^{2}}}.
\end{equation}
For $r_{M}\gg R$ (\ref{gen-sol}) transforms into
\begin{equation}\label{gen-sol-1}
\vec\phi\approx-\sqrt{f}\frac{\left(\frac{M\vec
l}{l^{3}}+\frac{m(r)\vec r}{r^{3}}\right)}{\sqrt{\left(\frac{M\vec
l}{l^{3}}+\frac{m(r)\vec r}{r^{3}}\right)^{2}}}.
\end{equation}

We can estimate the maximal value of the field $\phi$ itself. The
approximate size of the non-linearity zone is $r_{M}$, and $\phi$
can be estimated as $$\phi\sim
\sqrt{(\vec\phi\vec\phi)}\,r_{M}=\sqrt{f}r_{M}=\frac{M}{8\pi\sqrt{\omega}M_{Pl}r_{M}}$$
and thus
$$\frac{\phi}{\sqrt{\omega}M_{Pl}}\sim \frac{M}{8\pi\omega
M_{Pl}^{2}r_{M}}.$$ A more accurate analysis based on the use of
the solutions inside and outside the non-linearity zone (the
latter behaves as $\sim 1/L$, where $L\gg r_{M}$ is a
characteristic distance from both bodies), provides an analogous
estimate (up to the factor of the order of unity). For the
parameters, which will be considered in the next section,
$\frac{\phi}{\sqrt{\omega}M_{Pl}}\ll 1$ and the corresponding
terms in (\ref{act-BD22}) indeed can be omitted.

Now let us estimate the effects that could be produced by the
omitted term $-f$ of action (\ref{act-BD}) (see previous section).
In the non-linearity zone
$\sqrt{1-\frac{1}{f}({\vec\phi}{\vec\phi})}\ll 1$ and the term
$-f$ is not compensated. It indicates that the background metric
in the non-linearity zone is not the flat Minkowski metric, but a
de Sitter-like background metric leading to a local expansion. For
the observer, say, on the first body it looks like a repulsive
force acting on the second body. This force has the form
\begin{equation}
\vec F_{rep}\sim m\frac{f}{M_{Pl}^{2}}\vec R,
\end{equation}
which can be easily obtained by considering de Sitter metric in
the static form \cite{Weinberg}. Using (\ref{fexp}) we get
\begin{equation}
\left|\vec F_{rep}\right|\sim \left(\frac{M}{4\pi
M_{Pl}^{2}r_{M}}\right)\left(\frac{R}{r_{M}}\right)\frac{1}{16\pi
M_{Pl}^{2}}\frac{M}{\omega r_{M}^{2}}m.
\end{equation}
For the values of the parameters that will be used below this
force appears to be much smaller than the forces caused by the
fields $b_{\mu\nu}$ and $\phi$ obeying (\ref{eq-gr-b}) and
(\ref{eq1}) respectively. Thus, for our purposes we can use the
Minkowski background metric instead of a de Sitter-like background
metric. We would like to note that analogous estimates can be
obtained if we retain the term $-f$ in the action and get slightly
modified equations for the fields $b_{\mu\nu}$ and $\phi$.

\section{Specific examples}

Now we turn to the effects which can be produced by the DBI scalar
field in our Solar System. Let us suppose that
$$M=M_{\odot},\quad r_{M}\approx 100\, AU,\quad \omega\approx 700,$$
which means that $f\approx 2\cdot 10^{-44}\,GeV^{4}$ (our
"reduced" Planck mass $M_{Pl}\approx \frac{1.2\cdot
10^{19}GeV}{\sqrt{16\pi}}\approx 1.7\cdot 10^{18} GeV$).

In what follows we will consider two cases:
\begin{enumerate}
\item A light body with the mass $m$, for which
$\frac{m}{r_{*}^{2}}\sim \frac{m(r)}{r^{2}}\ll\frac{M}{R^{2}}$,
$r_{*}\ll R\ll r_{M}$ (for example, a spacecraft like Pioneer 10,
11 with $m\sim 300\, kg$, $r_{*}\sim 1\,m$). In this case in the
leading order
\begin{eqnarray}\label{19}
\vec\phi=-\left(\frac{M\vec l}{l^{3}}+\frac{m(r)\vec
r}{r^{3}}\right)\frac{\sqrt{f}}{\sqrt{\left(\frac{M\vec
l}{l^{3}}+\frac{m(r)\vec r}{r^{3}}\right)^{2}}}\approx\\
\nonumber -\left(\frac{M\vec l}{l^{3}}+\frac{m(r)\vec
r}{r^{3}}\right)\frac{\sqrt{f}}{\sqrt{\left(\frac{M}{l^{2}}\right)^{2}}}=-\sqrt{f}\left(\frac{\vec
R}{R}+\frac{m(r)\vec r R^{2}}{M r^{3}}\right).
\end{eqnarray}

It is worth mentioning that there is no such static solution for
the case $f<0$. Indeed, if $f<0$ then $r_{M}^{4}<0$ and we get
negative values under the square root (see Eq. (\ref{gen-sol})),
which is the consequence of the existence of a horizon at a finite
distance (see, for example, \cite{Nastase:2005pb}, where solutions
with horizons in DBI scalar field theory are discussed). That is
why we chose the case $f>0$.

Substituting the latter formula into (\ref{force}) and integrating
over the volume of the body leads to (we use the fact that
$\int\vec r d\Omega=0$, where $\Omega$ is the solid angle)
\begin{equation}\label{force-1}
{\vec F}_{anom}=-\frac{1}{16\pi M_{Pl}^{2}}\frac{M}{\omega
r_{M}^{2}}m.
\end{equation}
It is necessary to note that in principle
\begin{equation}\label{neq}
\int\frac{\vec r}{\sqrt{\left(\frac{M\vec l}{l^{3}}+\frac{m(r)\vec
r}{r^{3}}\right)^{2}}}\, d\Omega\ne 0.
\end{equation}
But we can neglect possible corrections because anyway
$$\left|\frac{M\vec l}{l^{3}}\right|\gg\left|\frac{m(r)\vec r}{r^{3}}\right|,$$
see (\ref{19}).

Formula (\ref{force-1}) is written in the system of units
$\hbar=c=1$. The replacement $\frac{1}{16\pi M_{Pl}^{2}}\to G$
allows one to pass to the SI units, which results in the
acceleration towards the Sun in the SI units
\begin{equation}
a_{anom}=\frac{GM}{\omega r_{M}^{2}}\approx 8.7\cdot
10^{-10}m/s^{2}.
\end{equation}
We note that this acceleration does not depend on the distance
from the Sun, which is exactly the situation with the Pioneer~10
and Pioneer~11 spacecraft \cite{Anderson:2001sg,Toth:2006qb}.

As for the bodies on the surface of the Earth, we can carry out
analogous calculations taking $M=M_{\oplus}$ (in this case $r_{M}$
also changes). Our ideal test bodies with density profile
(\ref{dens}), (\ref{dens1}) on the surface of the Earth also
possess an additional acceleration $a_{anom}$ towards the center
of the planet. It is evident that this acceleration can be
neglected in comparison with $g\approx 9.8\,m/s^{2}$ for
Earth-based gravitational experiments.
\item Heavy bodies with the mass $m$ (planets),
$\frac{M}{R^{2}}\ll\frac{m(r)}{r^{2}}$, $r_{*}\ll R\ll r_{M}$. In
this case we should carry out calculations more precisely because
$$\left|\frac{M\vec l}{l^{3}}\right|\ll\left|\frac{m(r)\vec r}{r^{3}}\right|,$$
and possible corrections due to (\ref{neq}) can be quite large.
Correspondingly, we should take
\begin{eqnarray}
\sqrt{\left(\frac{M\vec l}{l^{3}}+\frac{m(r)\vec
r}{r^{3}}\right)^{2}}\approx \frac{m(r)}{r^{2}}\sqrt{1+2\frac{M
r(\vec r\vec R)}{m(r)R^{3}}}\approx
\frac{m(r)}{r^{2}}\left(1+\frac{M r(\vec r\vec
R)}{m(r)R^{3}}\right)
\end{eqnarray}
and thus
\begin{equation}
-\frac{m(r)\vec r}{r^{3}}\frac{\sqrt{f}}{\sqrt{\left(\frac{M\vec
l}{l^{3}}+\frac{m(r)\vec r}{r^{3}}\right)^{2}}}\approx
-\sqrt{f}\left(\frac{\vec r}{r}-\frac{Mr^{2}}{m(r)R^{2}}\frac{\vec
r}{r}\left(\frac{\vec R}{R}\frac{\vec r}{r}\right)\right).
\end{equation}
Finally we obtain
\begin{eqnarray}\label{fin-eq}
\vec\phi=-\left(\frac{M\vec l}{l^{3}}+\frac{m(r)\vec
r}{r^{3}}\right)\frac{\sqrt{f}}{\sqrt{\left(\frac{M\vec
l}{l^{3}}+\frac{m(r)\vec r}{r^{3}}\right)^{2}}}\approx\\
\nonumber-\sqrt{f}\left(\frac{\vec
r}{r}+\frac{Mr^{2}}{m(r)R^{2}}\left[\frac{\vec R}{R}-\frac{\vec
r}{r}\left(\frac{\vec R}{R}\frac{\vec r}{r}\right)\right]\right).
\end{eqnarray}
Substituting the latter formula into (\ref{force}) and integrating
over the volume of the body leads to
\begin{equation}
{\vec F}_{anom}=-\frac{1}{16\pi M_{Pl}^{2}}\frac{8\pi M}{3\omega
r_{M}^{2}m}\left(\int_{0}^{r_{*}}\frac{\rho(r)}{m(r)}r^{4}dr\right)\frac{Mm\vec
R}{R^{3}},
\end{equation}
in the SI units
\begin{equation}\label{force-2}
{\vec F}_{anom}=-G_{eff}\frac{Mm\vec R}{R^{3}}
\end{equation}
where
\begin{equation}\label{geff}
G_{eff}=G\frac{8\pi M}{3\omega
r_{M}^{2}m}\int_{0}^{r_{*}}\frac{\rho(r)}{m(r)}r^{4}dr.
\end{equation}
One can see that the extra force acting on a heavy body $\sim
1/R^{2}$. Such a behavior is inherent to the ordinary Brans-Dicke
theory and we can replace the original potential $\phi$ in
(\ref{force}) by an effective potential $\sim 1/R$. The effective
Brans-Dicke parameter can be easily extracted from (\ref{geff}):
\begin{equation}\label{ombd1}
\frac{1}{2\omega_{BD}+3}=\frac{8\pi M}{3\omega
r_{M}^{2}m}\int_{0}^{r_{*}}\frac{\rho(r)}{m(r)}r^{4}dr,
\end{equation}
\begin{equation}
\omega_{BD}\approx\frac{3\omega r_{M}^{2}m}{16\pi
M\int_{0}^{r_{*}}\frac{\rho(r)}{m(r)}r^{4}dr}.
\end{equation}
A significant difference from the original Brans-Dicke theory is
that $\omega_{BD}$ depends on the mass $m$, i.e. it is different
for different planets.

To estimate $\omega_{BD}$ for different planets we suppose that
$\rho(r)=\frac{3m}{4\pi r_{*}^{3}}=const$. In this case
\begin{equation}\label{ombd2}
\omega_{BD}\approx\frac{\omega
mr_{M}^{2}}{2Mr_{*}^{2}}=\frac{a_{ff}}{2a_{anom}},
\end{equation}
where $a_{ff}$ is the free fall acceleration on the surface of a
body (a planet). For example,
$$\omega_{BD}^{Mercury}\approx 2.1\cdot 10^{9},$$
$$\omega_{BD}^{Jupiter}\approx 1.4\cdot 10^{10}.$$
Such large values of the Brans-Dicke parameter do not contradict
the experimental bounds $\omega_{BD}>3500$ obtained in the Solar
System gravitational tests \cite{Will:2005va,Chiba:2003ir} (we
would like to note that the bounds on the $\sim {\vec R}/R^{3}$
extra force differ from the bounds on the $\sim {\vec R}/R$ extra
force).

For the other limiting case $\rho(r)=\frac{m}{2\pi r_{*}^{2}r}$
\begin{equation}
\omega_{BD}\approx\frac{3a_{ff}}{4a_{anom}}.
\end{equation}

One should note that in the case of a heavy body with
$\rho(r)=const$ there exists a region $r<\hat r$ such that
$\frac{M}{R^{2}}\approx \frac{m(\hat r)}{{\hat r}^{2}}$. For this
region one should carry out the calculations described in item~1
(the case of a light body). But even for Mercury (if we suppose
$\rho(r)=const$) $\frac{\hat r}{r_{*}}\approx 1.8\cdot 10^{-2}$,
and the constant extra acceleration appears to be
$$a_{extra}\approx \frac{\hat r^{3}}{r_{*}^{3}}\,a_{anom} \approx 6\cdot 10^{-6}a_{anom},$$
which does not contradict the existing experimental restriction on
a possible extra constant acceleration of the planet
\cite{Sanders:2006yf} (which should be much smaller than that of
the Pioneers 10, 11 spacecraft). It is easy to check that for the
other planets of the Solar System $a_{extra}$ also does not exceed
the experimentally allowed limits \cite{Sanders:2006yf}. If one
takes $\rho\sim\frac{1}{r}$ this region is absent and thus
$a_{extra}=0$. Of course, our density profiles for the planets are
an idealization, but more realistic profiles should lead to the
values of $a_{extra}$ which lie somewhere between the limiting
values obtained above. The same is valid for the values of the
effective Brans-Dicke parameter $\omega_{BD}$.
\end{enumerate}
The physical difference between the two cases can be easily
explained. In the first case vectors $\vec\phi$ are directed
approximately parallel to the vector $\vec R$ at any point of the
second body, whereas in the second case these vectors are
approximately parallel to the radius-vectors $\vec r$, which leads
to the result discussed above.

Thus we have shown that in principle it is possible to construct a
covariant theory which "distinguishes" light and heavy test bodies
with respect to an external gravitational field of a source. We
would also like to note that the value of the parameter $f$ which
is necessary to reproduce the anomalous acceleration of the
Pioneer spacecraft was chosen to be $f\approx 2\cdot
10^{-44}\,GeV^{4}$. This value is quite close to the vacuum energy
density $\sim 10^{-47}\,GeV^{4}$, responsible for the accelerating
expansion of the Universe. In this connection it is very
interesting to examine possible cosmological manifestations of the
model described by action (\ref{act-BD}). This issue calls for a
more detailed and thorough investigation.

Of course we can not argue that the Pioneer anomaly is indeed
caused by the existence of such a DBI scalar field. Moreover,
recently it was shown that a part of the anomalous acceleration
can be explained by the thermal recoil force effect
\cite{Toth:2009se}. Nevertheless the model presented in this paper
possesses quite interesting features, does not contradict
experimental data at least in the Newtonian limit and seems to be
worth an additional examination.

\section*{Acknowledgments}
The author is grateful to D.G. Levkov, O.V. Pavlovsky, M.S.
Pshirkov and I.P. Volobuev for valuable discussions. The work was
supported by grant for young scientists MK-5602.2008.2 of the
President of Russian Federation, grant of the "Dynasty"
Foundation, grant of Russian Ministry of Education and Science
NS-1456.2008.2, FASI state contract 02.740.11.0244, RFBR grant
08-02-92499-CNRSL$\_$a and scholarship for young teachers and
scientists of M.V. Lomonosov Moscow State University.

\end{document}